\documentclass[aps,prd,twocolumn,showpacs,showkeys,nofootinbib]{revtex4-1}

\usepackage{amsmath}
\usepackage{amssymb}
\usepackage{amsfonts}
\usepackage{hyperref}
\usepackage[]{units}
\usepackage{graphicx}
\usepackage{bm}

\DeclareMathOperator{\Tr}{Tr}
\usepackage{color}
\definecolor{change}{rgb}{0.0, 0.0, 0.0} 
\definecolor{change2}{rgb}{0.0, 0.0, 0.0} 

\begin{document}


\title{Holographic Entanglement Entropy in the QCD Phase Diagram with a Critical Point}


\author{J. Knaute}
\email{j.knaute@hzdr.de}

\author{B. K\"ampfer}

\affiliation{Helmholtz-Zentrum Dresden-Rossendorf, POB 51 01 19, 01314 Dresden, Germany}
\affiliation{TU Dresden, Institut f\"ur Theoretische Physik, 01062 Dresden, Germany}



\begin{abstract}
We calculate the holographic entanglement entropy for the holographic QCD phase diagram considered in \cite{Knaute:2017opk} and explore the resulting qualitative behavior over the temperature-chemical potential plane.  
In agreement with the thermodynamic result, the phase diagram exhibits the same critical point 
as the onset of a first-order phase transition curve.
We compare the phase diagram of the entanglement entropy to that of the thermodynamic entropy density and 
find a striking agreement in the vicinity of the critical point.
Thus, the holographic entanglement entropy qualifies to characterize different phase structures.
The scaling behavior near the critical point is analyzed through the calculation of critical exponents.
\end{abstract}


\maketitle

\section{Introduction}

The AdS/CFT correspondence \cite{Maldacena:1997re,Gubser:1998bc,Witten:1998qj} or more general gauge/gravity duality provides a helpful tool to explore properties of strong-coupling systems and in particular the QCD phase diagram. 
In \cite{Knaute:2017opk} a holographic QCD phase diagram was presented, which is adjusted to 2+1 flavor lattice QCD with physical quark masses \cite{Borsanyi:2013bia,Bazavov:2014pvz,Bellwied:2015lba} 
and results in a critical endpoint (CEP) at a temperature $T_{CEP}\approx\unit[112]{MeV}$ and a baryo-chemical potential $\mu_{CEP}\approx\unit[612]{MeV}$ as the starting point of a first-order phase transition (FOPT) curve
towards larger chemical potential. 
The setup for this bottom-up approach was originally formulated in  \cite{DeWolfe:2010he,DeWolfe:2011ts} and further investigated, e.g., in \cite{Rougemont:2015wca,Rougemont:2015ona,Rougemont:2017tlu}. 

Beyond thermodynamic quantities also non-local observables such as entanglement entropy play an important role. 
Entanglement entropy is used extensively to characterize phases, as an order parameter for phase transitions and as a measure of degrees of freedom or quantum information in physical systems.
(See e.g.\ \cite{Terashima:1999vw,Vidal:2002rm,Calabrese:2004eu,OrusLacort:2006ni,Calabrese:2009qy,Rovelli:2011ur,Laflorencie:2015eck} and references therein for a small but interesting selection of different  topics.)
A holographic formula for this quantity was proposed in \cite{Ryu:2006bv,Ryu:2006ef} as the minimal surface in the bulk for a given boundary. 
(See \cite{Nishioka:2009un,Rangamani:2016dms} for reviews on that topic.)
This concept has attracted enormous attention to study the Van der Waals-like phase transition in charged Reissner-Nordstr\"om-AdS black holes \cite{Chaturvedi:2016kbk,Li:2017gyc,Zeng:2016fsb} 
and massive \cite{Zeng:2015tfj} or Weyl \cite{Dey:2015ytd} gravity. 
Moreover, it was analyzed to characterize thermalization processes \cite{Caceres:2012em,Ageev:2017wet}, 
and in the context of the gravity/condensed matter correspondence \cite{Takayanagi:2013xqa}~- 
particularly in studies of holographic superconductors \cite{Albash:2012pd,Cai:2012sk,Cai:2012nm,Johnson:2013dka,Dey:2014voa,Zangeneh:2017tub} 
and metal-insulator transitions \cite{Ling:2015dma,Ling:2016wyr,Ling:2016dck}.
Very recently, an experimental attempt to measure holographic entanglement entropy (HEE) on a quantum simulator in the context of tensor networks was presented \cite{Li:2017qwu}. 
Holographic entanglement entropy might thus provide a promising approach to study and verify quantum gravity effects in realistic systems and experiments.

In \cite{Klebanov:2007ws} it was first discussed that HEE can serve as a probe of confinement in gravity duals of large-$N_c$ gauge theories: The change between connected and disconnected surfaces in dependence of the length of the boundary area was interpreted as a signature of confinement. 
(Further investigations on that topic can be found, e.g., in \cite{Lewkowycz:2012mw,Kim:2013ysa,Kol:2014nqa,Ghodrati:2015rta,Dudal:2016joz}.) 
This confinement-deconfinement transition of entanglement entropy in non-Abelian gauge theories was also studied on the lattice \cite{Velytsky:2008rs,Buividovich:2008kq,Itou:2015cyu}.
Recently, a discussion on entanglement entropy in strongly coupled systems was presented \cite{Zhang:2016rcm}:
It was discussed that the behavior of entanglement entropy can characterize different phase structures in a holographic model proposed in \cite{Gubser:2008ny,Gubser:2008yx}. 
The main difference to the previous analyses mentioned above is the discussion in dependence on the temperature for a fixed boundary configuration. 
Here, we extend these studies for the holographic QCD model in \cite{Knaute:2017opk} in dependence on the temperature and chemical potential. 
(See also \cite{Kundu:2016dyk} for some aspects on the behavior of HEE in Reissner-Nordstr\"om geometries at finite chemical potential.)

\section{Review of the holographic EMd model}

The holographic QCD phase diagram at finite temperature and chemical potential in \cite{Knaute:2017opk} is based on a Einstein-Maxwell-dilaton (EMd) model which was initially formulated in \cite{DeWolfe:2010he}. We refer to these references for details and present here just a very brief summary of the setup. 

The defining action is
 \begin{equation} \label{eqn:S}
S = \frac{1}{2 \kappa_{5}^2}\int d^{5}x\sqrt{-g}\left(R-\frac{1}{2}\partial^{\mu}\phi\partial_{\mu}\phi-V(\phi) - \frac{f(\phi)}{4}F^2_{\mu\nu}\right), 
\end{equation}
where $F_{\mu\nu} = \partial_\mu A_{\nu} - \partial_\nu A_{\mu}$ with $A_\mu dx^\mu = \Phi dt$ is the Abelian gauge field, $V(\phi)$ stands for the potential describing the self-interaction of the dilaton $\phi$, $f(\phi)$ is a dynamical strength function that couples the dilaton and gauge field, and $\kappa_5$ is the 5-dimensional gravitational constant. 
The metric ansatz
\begin{equation} \label{eqn:ds2}
ds^2 = \mathrm e^{2 A(r)} \left( -h(r) dt^2 + d\vec x^2 \right) + \frac{dr^2}{h(r)} 
\end{equation}
represents an asymptotically AdS$_5$ spacetime with boundary at $r\to\infty$ and defines a black hole horizon by $h(r_H) \equiv 0$. 
The field equations following from (\ref{eqn:S},\,\ref{eqn:ds2}) are solved 
{\color{change}numerically (cf.\ \cite{Knaute:2017opk,DeWolfe:2010he} for technical aspects) for the metric coefficients $h(r)$ and $A(r)$ as well as the profiles $\Phi(r)$ and $\phi(r)$} 
with $\phi_0 \equiv \phi(r_H)$ and $\Phi_1 \equiv \frac{\partial \Phi}{\partial r}\big\vert_{r_H}$ as the only remaining independent parameters, {\color{change}which serve as initial conditions}.
The thermodynamic quantities temperature $T$, entropy density $s$, baryo-chemical potential $\mu$ and baryon 
density $n$ are then calculated using the boundary expansions of the {\color{change}such obtained} functions $h(r)$, $A(r)$, $\Phi(r)$ and $\phi(r)$. 
In \cite{Knaute:2017opk}, multi-parameter ans\"atze for the potential $V(\phi)$ and gauge kinetic function $f(\phi)$ were elaborated that mimic the QCD equation of state (EoS) and second-order quark number susceptibility of the 2+1 flavor lattice QCD data {\color{change}with physical quark masses} \cite{Borsanyi:2013bia,Bazavov:2014pvz,Bellwied:2015lba} at $\mu=0$ very precisely.\footnote{In \cite{Borsanyi:2016ksw}, results for 3+1 flavor lattice QCD have been presented. Since charm quarks impact to the EoS only for temperatures above \unit[250]{MeV}, our holographic model still allows a good description in the relevant temperature region of the CEP.} 
{\color{change}The explicit forms of these functions as well as further details of the EMd model are discussed in Appendix \ref{app:details}.}
The $T\negmedspace-\negmedspace\mu$ plane is then uncovered within the framework of this EMd model by properly chosen initial conditions 
$(\phi_0,\Phi_1)$.

\section{Holographic Entanglement Entropy}

Consider a quantum mechanical system which is 
(i) described by the density operator $\rho_{tot}$ and 
(ii) divided into a subsystem $\mathcal A$ and its complement $\mathcal B$. 
The entanglement entropy of $\mathcal A$ is defined as the von Neumann entropy 
\begin{equation} \label{eqn:EE}
S_{\text{EE}} := -\Tr_{\mathcal A} \rho_{\mathcal A} \ln \rho_{\mathcal A}
\end{equation}
w.r.t.\ the reduced density matrix $\rho_{\mathcal A} = \Tr_{\mathcal B} \rho_{tot}$.
According to \cite{Ryu:2006bv,Ryu:2006ef}, the holographic dual of this quantity for a CFT$_d$ on $\mathbb R^{1,d-1}$ is given as
\begin{equation} \label{eqn:HEE}
S_{\text{HEE}} = \frac{\text{Area}(\gamma_{\mathcal A})}{4 G_N^{(d+1)}}  ,
\end{equation}
where $\gamma_{\mathcal A}$ is the static minimal surface in AdS$_{d+1}$ with boundary $\partial \gamma_{\mathcal A}=\partial \mathcal A$ and $G_N^{(d+1)}$ is the $d+1$ dimensional Newton constant. 
In the present work, we analyze the behavior of entanglement entropy in the holographic QCD phase diagram \cite{Knaute:2017opk} near the critical point.
Similar to \cite{Zhang:2016rcm}, we assume a fixed strip shape on the boundary for the entanglement region
\begin{equation}
\mathcal A: \quad x_1 \in [-l/2,l/2] , \quad x_2,x_3 \in [-L/2,L/2]
\end{equation}
with $L \gg l$ such that translation invariance is preserved and the minimal surface can be parameterized by the single function $r=r(x_1)$.
The induced metric on the static minimal surface is
\begin{equation}
ds^2_{\gamma_{\mathcal A}} = \left( \mathrm e^{2A}+\frac{r^{\prime 2}}{h} \right) dx_1^2 + \mathrm e^{2A} \left( dx_2^2 + dx_3^2 \right) ,
\end{equation}
where a prime denotes a derivative w.r.t.\ $x_1$.
The HEE \eqref{eqn:HEE} then follows as
\begin{align}
S_{\text{HEE}} &= \frac{1}{4} \int dx_1 dx_2 dx_3 \sqrt{\gamma} \\
                        &= \frac{V_2}{2} \int_0^{l/2} dx_1 \mathrm e^{2A(r)} \sqrt{\mathrm e^{2A(r)}+\frac{r^{\prime 2}}{h(r)}}
\label{eqn:SHEE_v2}                   
\end{align}
with $\gamma$ as the determinant of the induced metric on $\gamma_{\mathcal A}$ and $V_2\equiv \int dx_2 dx_3$.
Extremizing $S_{\text{HEE}}$ by taking into account conserved quantities, one finds
\begin{align}
\sqrt{\mathrm e^{2A(r)}+\frac{r^{\prime 2}}{h(r)}} &= \frac{\mathrm e^{4A(r)}}{\mathrm e^{3A(r_*)}} \label{eqn:conserved} \\
\Longleftrightarrow\quad r^\prime &= \sqrt{h(r)\left( \mathrm e^{8A(r)-6A(r_*)} - \mathrm e^{2A(r)} \right)} \label{eqn:dr} , 
\end{align}
where $r_*$ is the closest position of the minimal surface to the horizon.  
Integrating Eq.\,\eqref{eqn:dr} w.r.t.\ the boundary condition
\begin{equation}
\frac{l}{2} = \int_{r_*}^\infty dr \left[ h(r)\left( \mathrm e^{8A(r)-6A(r_*)} - \mathrm e^{2A(r)} \right) \right]^{-1/2} ,
\label{eqn:bc}
\end{equation}
one can solve Eq.\,\eqref{eqn:bc} for $r_*$ for a given $l$.
Then, $S_{\text{HEE}}$ follows by plugging \eqref{eqn:conserved} and \eqref{eqn:dr} into \eqref{eqn:SHEE_v2} as
\begin{align}
S_{\text{HEE}} &= \frac{V_2}{2} \int_0^{l/2} dx_1 \frac{\mathrm e^{6A(r)}}{\mathrm e^{3A(r_*)}} \\
                        &= \frac{V_2}{2} \int_{r_*}^\infty dr \frac{\mathrm e^{6A(r)-3A(r_*)}}{\mathrm e^{A(r)} \sqrt{h(r)  \left( \mathrm e^{6A(r)-6A(r_*)} -1 \right)}} . \label{eqn:SHEE_full}
\end{align}
This quantity is divergent. 
{\color{change}Desirable would be a systematic regularization and renormalization, e.g.\ by suitable counterterms, similarly to \cite{Taylor:2016aoi,Taylor:2017zzo}. We postpone such an intricate investigation in its own right to follow-up work and explore instead an ad-hoc}
regularized HEE density as
\begin{equation} \label{eqn:SHEE_reg}
S_{\text{HEE}}^{reg}  := \frac{1}{2} \int_{r_*}^{r_m} dr \frac{\mathrm e^{6A(r)-3A(r_*)}}{\mathrm e^{A(r)} \sqrt{h(r)  \left( \mathrm e^{6A(r)-6A(r_*)} -1 \right)}} , 
\end{equation}
where $r_m$ is a sufficiently large cutoff,  similarly to be employed in  Eq.\,\eqref{eqn:bc}. 

In addition, we consider also a renormalized HEE density by the following construction:
Denote the integrand in Eq.\,\eqref{eqn:SHEE_full} as $H(r)$ and define $\tilde H(r)$ by setting $A(r_*)\equiv0$ in $H(r)$. 
{\color{change}As shown in \cite{DeWolfe:2010he}, $h$ goes as $h(r) = h_0^\infty + \ldots$ like a constant for $r\to\infty$ at the boundary  and $A(r) = \frac{1}{\sqrt{h_0^\infty}} r + A_0^\infty + \dots$ is linear. 
The integrand $H(r)$ thus behaves like $\exp\left\{2r / \sqrt{h_0^\infty}\right\}$ for large $r$. Since the metric functions converge quickly to their asymptotic values, $H(r)$ diverges generically like $1/\sqrt{r}$ for small $r$, i.e.\ near the horizon.}
The function $\tilde H(r)$ has the same boundary asymptotics but deviates near $r_*$ {\color{change}and we want to consider the finite renormalized integrand $H(r)-\tilde H(r)$. Since the numerical values in this difference become very large, we turn to the logarithm and define a renormalized HEE density as} \footnote{Note that 
contrary to \cite{Zhang:2016rcm} we do not introduce a renormalized density w.r.t.\ some reference point, since this procedure yields negative values, which we do not interpret  as physical, because they are not possible in the original definition \eqref{eqn:EE}.}
\begin{equation} \label{eqn:SHEE_ren}
S_{\text{HEE}}^{ren}  := \frac{1}{2} \int_{r_*}^{r_m} dr \ln \frac{H(r)}{\tilde H(r)} .
\end{equation}
{\color{change2}In general, there is also the possibility of a disconnected entangling surface which reaches from the boundary at $r=\infty$ up to the horizon at $r_* = r_H = 0$. 
We postpone the consideration of such a surface class to separate investigations which require the extension of the present numerical apparatus. The latter one is here optimized for numerical solutions of the metric functions from (slightly) outside the horizon towards the boundary and does not include them.}

\section{Phase diagram}

\begin{figure}[!t]
\centering
 \includegraphics[width=\columnwidth]{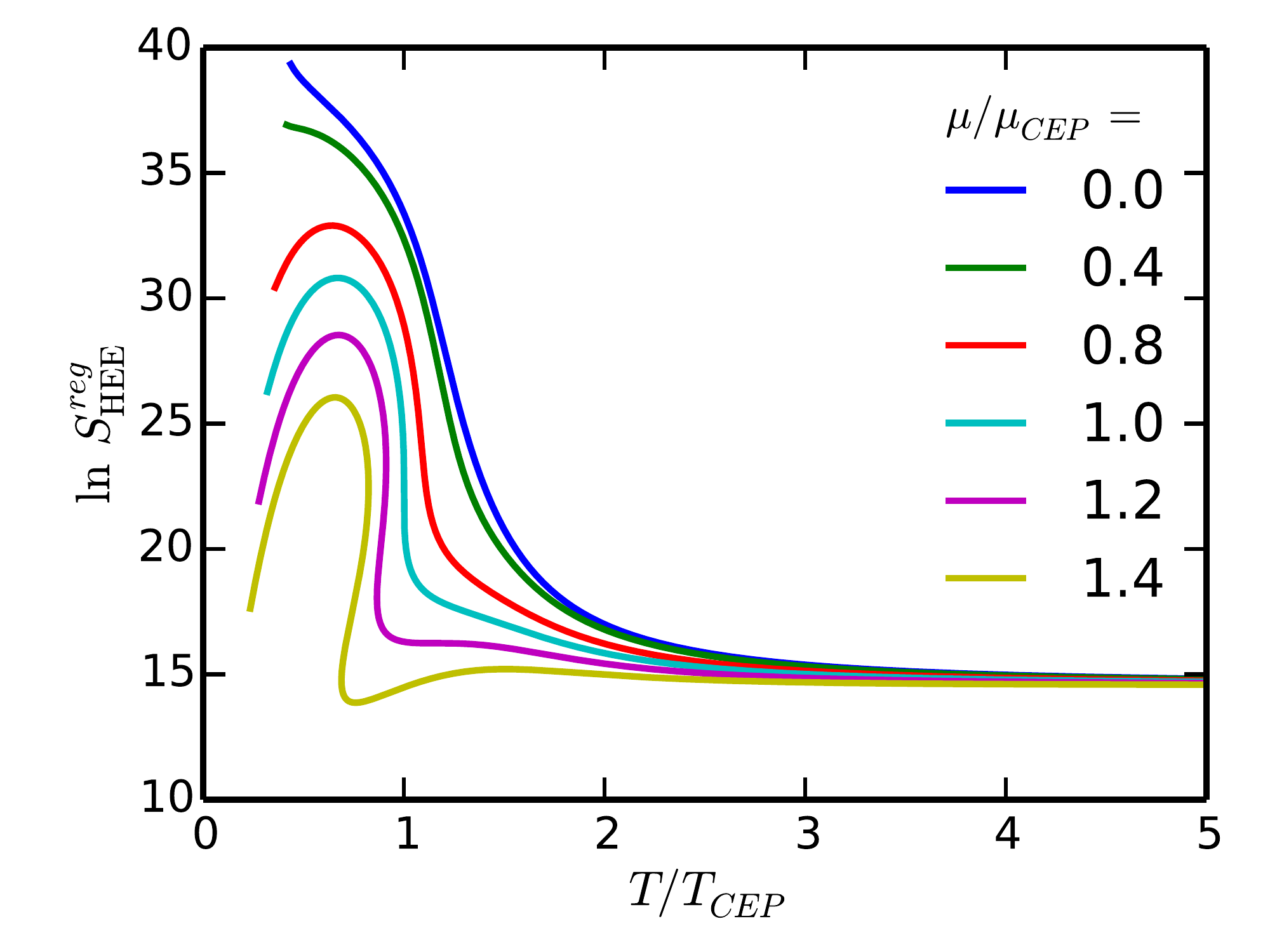}
 \caption{Regularized holographic entanglement entropy density $\ln S_\text{HEE}^{reg}$ as a function of the temperature $T$ for different values of the chemical potential $\mu$. }
 \label{fig:mulvls}
\end{figure}

\begin{figure*}[!t]
\centering
 \includegraphics[width=0.49\textwidth]{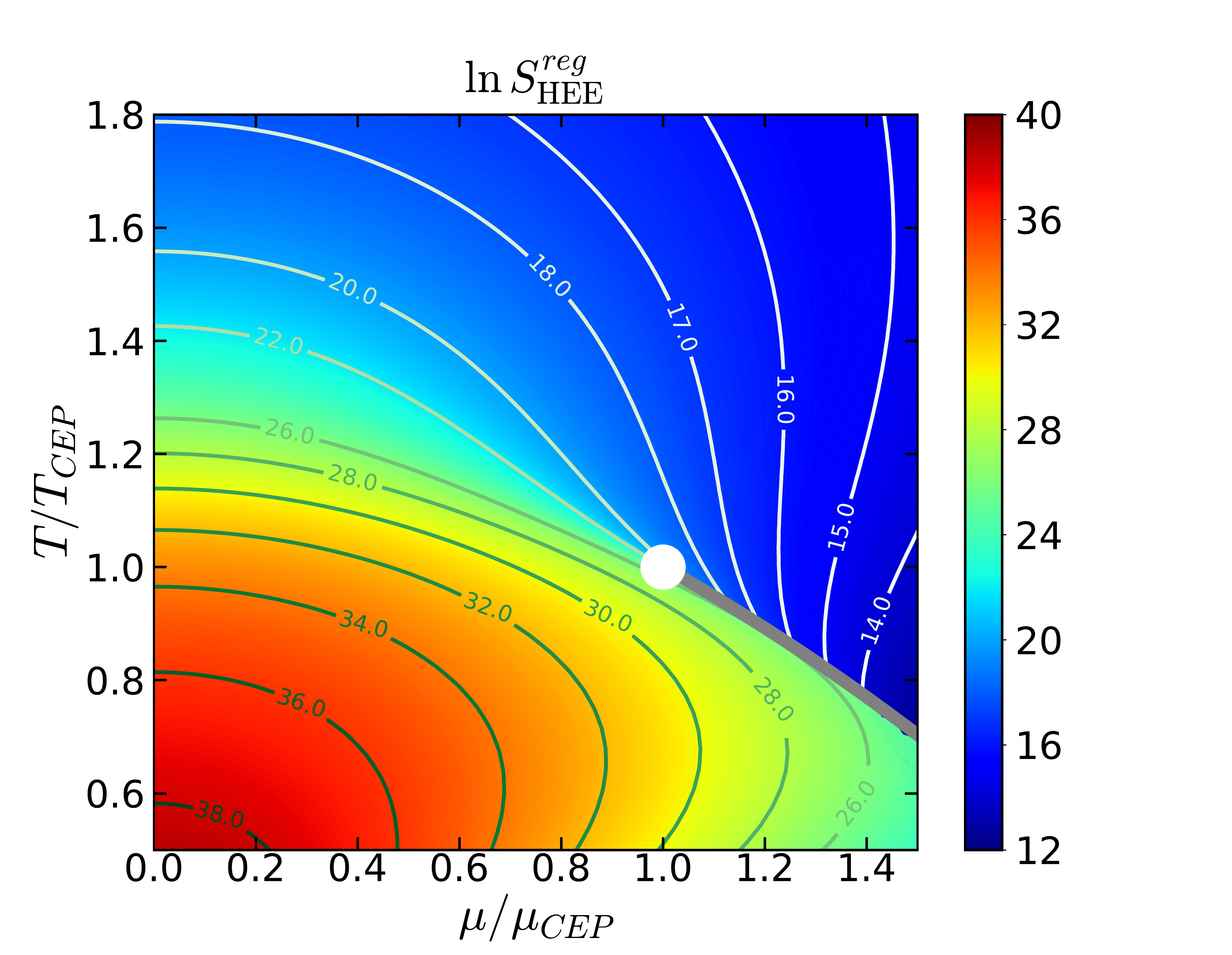}
 \includegraphics[width=0.49\textwidth]{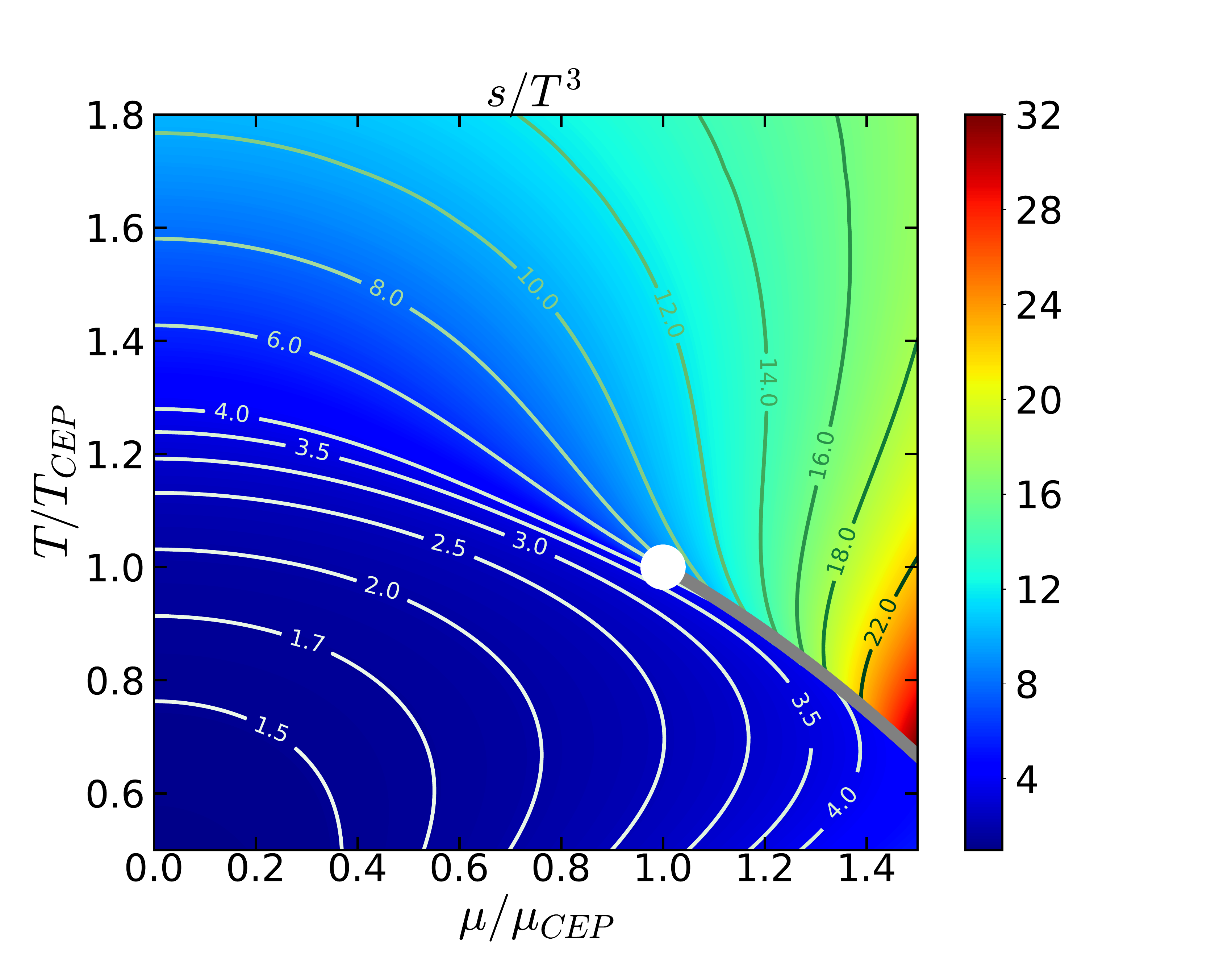}
 \caption{Contour plots of the regularized holographic entanglement entropy density $\ln S_\text{HEE}^{reg}$ (left) and scaled entropy density $s/T^3$ (right) over the $T\negmedspace-\negmedspace\mu$ plane. The position of the CEPs are marked by a white dot and the FOPT curves are displayed as grey lines.}
 \label{fig:PD}
\end{figure*}

\begin{figure}[!t]
\centering
 \includegraphics[width=\columnwidth]{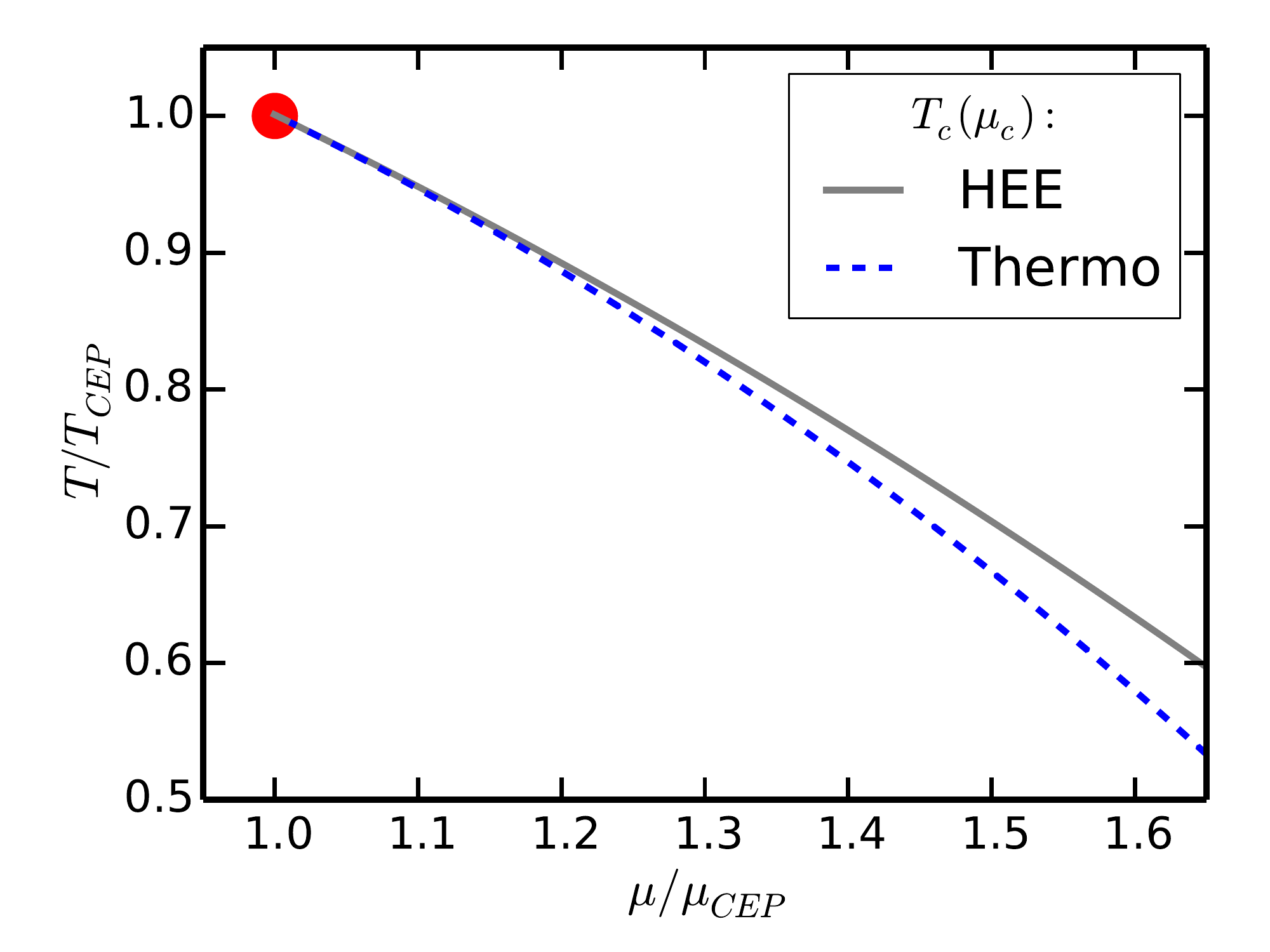}
 \caption{Comparison of FOPT curves over the $T\negmedspace-\negmedspace\mu$ plane based on the left panel of Fig.\,\ref{fig:PD} (grey curve) and the result exhibited in the right panel of Fig.\,\ref{fig:PD} (blue dashed curve). The position of the CEP is marked by a red dot.}
 \label{fig:FOPT}
\end{figure}

We calculated the HEE density \eqref{eqn:SHEE_reg} as outlined in the previous paragraph for numerically generated charged black hole solutions with initial conditions $\phi_0 \in [0.35, 4.5]$ and 
$\Phi_1/\Phi_1^{max}(\phi_0) \in [0, 0.755]$ as in \cite{Knaute:2017opk} and set the width of the entanglement strip to $l=0.04$. For the following qualitative study we choose $r_m=2.0$ and checked that the behavior is similar also for larger cutoff values. 

Figure \ref{fig:mulvls} shows $S_{\text{HEE}}^{reg}$ in dependence on the temperature for different values of the chemical potential. 
For $\mu=0$, $S_{\text{HEE}}^{reg}$ is monotonically decreasing in the characteristic crossover region $T = \mathcal{O}(150 \text{ MeV})$. 
The entanglement entropy is pushed towards smaller values with increasing chemical potential. 
A first-order phase transition at large values of $\mu$ is signaled by the appearance of a multivalued branch. 
{\color{change}($S_{\text{HEE}}^{ren}$ from \eqref{eqn:SHEE_ren} displays the same feature. This provides some confidence that both definitions - even being rather ad-hoc - yield robust results. Since \eqref{eqn:SHEE_ren} is numerically more demanding we continue to use \eqref{eqn:SHEE_reg}.)}
The asymptotically constant value of $S_{\text{HEE}}^{reg}$ at large $T$ is nearly independent of $\mu$. 
Since entanglement entropy can be interpreted as a measure for the \textit{quantumness} of a physical system, large values of $S_{\text{HEE}}^{reg}$ at small temperatures indicate the quantum region of the holographic QCD phase diagram, whereas the thermodynamic region at large $T$ and/or $\mu$ is characterized through a nearly constant entanglement entropy. 

Inspired by standard thermodynamic relations, we define a pseudo-pressure $p_\text{HEE}$ through the integration 
$dp_\text{HEE} = \ln(S_\text{HEE}^{reg}) \mathrm dT$ for $\mu \equiv \text{const}$, 
which exhibits an analogous pressure loop as in case of a FOPT and allows the definition of a transition temperature $T_c$. 

Figure \ref{fig:PD} shows the resulting phase diagram of the regularized HEE density over the $T\negmedspace-\negmedspace\mu$ plane (left panel). 
The CEP is located at $T_{CEP}=\unit[(111.5\pm0.5)]{MeV}$ and $\mu_{CEP}=\unit[(611.5\pm0.5)]{MeV}$ in agreement with the thermodynamic result of \cite{Knaute:2017opk}.
The stable phases of the HEE are discontinuous across the FOPT and jump towards smaller values with increasing temperature or chemical potential.

The right panel of Fig.\,\ref{fig:PD} shows the scaled standard thermodynamic-statistical entropy density $s/T^3$ over the $T\negmedspace-\negmedspace\mu$ plane for a comparison. 
The behavior of the thermodynamic entropy is opposite to the HEE, i.e.\ the entropy is increasing for larger values of $T$ or $\mu$ and jumps towards higher values across the FOPT, as typical for a gas-liquid transition. 
Despite these differences, the patterns of the scaled isentropes exhibit a remarkable similarity in both phase diagrams.\footnote{In fact,
the shape of the renormalized HEE density $S_{\text{HEE}}^{ren}$ in \eqref{eqn:SHEE_ren} resembles much
better $s/T^3$, as pointed out in \cite{Zhang:2016rcm} for vanishing $\mu$. 
Thus, $S_{\text{HEE}}^{ren}$ exhibits an opposite qualitative behavior, 
i.e.\ the decreasing behavior of $S_{\text{HEE}}^{reg}$ corresponds to an increasing behavior of 
$S_{\text{HEE}}^{ren}$ etc.
{\color{change}The mutual consistency of $S_{\text{HEE}}^{reg}$ and $S_{\text{HEE}}^{ren}$ w.r.t.\ the phase structure has been stressed already above.}} 

The exact locations of the FOPT  curves $T_c(\mu_c)$ are explicitly compared in Fig.\,\ref{fig:FOPT} based on the HEE pseudo-pressure definition and the true thermodynamic stability criterion. 
The two curves agree very well in the vicinity of the critical point up to $\mu_c/\mu_{CEP} \approx 1.2$ but deviate from each other approximately $5\,\%$ for $\mu_c/\mu_{CEP} \approx 1.6$.

\section{Critical Behavior}

Critical exponents describe the universal behavior of physical quantities near the critical point. 
Specifically, they quantify the divergence of derivatives of the free energy as power laws. 
Here, we are interested in the power law dependence of the specific heat at constant chemical potential:
\begin{equation}
C_\mu \equiv T \frac{\partial s}{\partial T}\Big\vert_\mu = -T \frac{\partial^2 f}{\partial T^2}\Big\vert_\mu
 \sim \vert T - T_{CEP} \vert^{-\alpha} , 
\end{equation}
where $\mu=\mu_{CEP}$ and $T<T_{CEP}$ are assumed. 
A similar definition holds for $\alpha^\prime$, where the critical point is approached for $T>T_{CEP}$.\footnote{Note 
that the critical exponent $\alpha$ for $C_n$, i.e.\ the heat capacity at constant baryon density along the FOPT curve, has the mean field result $\alpha=\alpha^\prime=0$.}
To determine $\alpha$, we consider the dependence $\vert T-T_{CEP}\vert \sim \vert s-s_{CEP}\vert^\beta$
and calculate $\beta$ through the linear fit function $\ln \vert T-T_{CEP}\vert = \beta \ln \vert s-s_{CEP}\vert + \text{const}$.
The critical exponent then follows as $\alpha=1-1/\beta$.
This procedure yields the following results for the thermodynamic entropy:
\begin{equation}
\alpha \approx 0.66 , \quad \alpha^\prime \approx 0.64 .
\end{equation}
For the HEE, we employ the logarithmic values $\ln S_\text{HEE}^{reg}$ and find
\begin{equation}
\alpha_{\text{HEE}} \approx 0.65 , \quad \alpha^\prime_{\text{HEE}} \approx 0.66 .
\end{equation}
Both results for the critical exponents yield nearly the same values for the second-order phase transition 
and agree well with the {\color{change}van der Waals criticality in AdS black holes \cite{Bhattacharya:2017nru}} $\alpha = \alpha^\prime = 2/3$.

\section{Discussion and Summary}

In the present note we study the qualitative behavior of the holographic entanglement entropy (HEE) in the holographic QCD phase diagram of \cite{Knaute:2017opk}. The setup rests on a Einstein-Maxwell-dilaton model \cite{DeWolfe:2010he,DeWolfe:2011ts} which was adjusted in \cite{Knaute:2017opk} to 2+1 flavor lattice QCD data with physical quark masses \cite{Borsanyi:2013bia,Bazavov:2014pvz,Bellwied:2015lba} to reproduce the QCD equation of state and quark number susceptibility. 

Here we explore the phase structure of the HEE over the temperature-chemical potential plane by introducing a cutoff to regularize the divergent entropy integral. 
A first-order phase transition (FOPT) is setting in at a critical endpoint (CEP) 
consistent with the result in \cite{Knaute:2017opk}. 
{\color{change}This is supported quantitatively also by another ad-hoc definition of a renormalized HEE.}
The precise course of the FOPT curve is determined by the definition of a pseudo-pressure as an integral over the HEE density. 
The resulting HEE FOPT curve agrees astonishing well with the FOPT curve based on the  thermodynamic stability criterion in the vicinity of the CEP.

The behavior of the regularized HEE density is opposite to the thermodynamic entropy: In the crossover region of the phase diagram, the HEE drops rapidly as a function of the temperature and jumps towards smaller values across the FOPT curve. This behavior separates the quantum region of the phase diagram from the region of dominating thermal fluctuations.

The logarithmic values of the regularized HEE density show a similar scaling behavior near the critical point as the thermodynamic entropy density. 
The critical exponents of the heat capacity at constant chemical potential agree well with the {\color{change}van der Waals criticality}.

These results indicate that HEE is capable of characterizing the different phases in the holographic QCD phase diagram, in particular in the vicinity of the CEP and the confinement/deconfinement transition. However, the HEE alone does not provide enough information to calculate the exact thermodynamic FOPT curve 
and the qualitative behavior depends on whether a regularization or renormalization scheme is applied.\\

Acknowledgements: We thank S.-J.\ Zhang for communications on holographic entanglement entropy.

\appendix
{\color{change}
\section{Details of the holographic EMd model}
\label{app:details}

The explicit forms of the dilaton potential and dynamical strength function in \cite{Knaute:2017opk} are
\begin{align}
L^2 V(\phi) &= N(\phi) \exp\left\{ \sum_{i=1}^4 a_i \phi^i + a_5\tanh\left[ a_6(\phi-\phi_a) \right] \right\} ,\\
N(\phi) &= b_0 + b_1\cosh^{b_3}\left[ b_2(\phi-\phi_b) \right] ,\\
f(\phi) &= c_0 + c_1\tanh\left[ c_2(\phi-\phi_c)\right] + c_3 \exp\left[-c_4 \phi\right]
\end{align}
with coefficients
\begin{widetext}
\begin{equation}
\begin{tabular}{c|c|c|c|c|c|c||c|c|c|c}
&$a_1$ & $a_2$ & $a_3$ & $a_4$ & $a_5$ & $a_6$ & $b_0$ & $b_1$ & $b_2$ & $b_3$ \\ \hline 
$\phi < \phi_m$ &  0 & 0.1420 & 0 & -0.0022 & 0 & 0 & -12 & 0 & 0 & 0 \\ \hline
$\phi \geq \phi_m$ & -0.0113 & 0 & 0 & 0 & -0.2195 & 2.1420 & 0 & -10.0138 & 0.4951 & 1.4270
\end{tabular} 
\end{equation}
\end{widetext}
and
\begin{equation}
\begin{split}
\begin{tabular}{c|c|c|c}
$\phi_m$ & $\phi_a$ & $\phi_b$ & $\phi_c$ \\ \hline 
1.7058 & 4.3150 & 0.1761 & 2.1820
\end{tabular} , \\
\begin{tabular}{c|c|c|c|c}
$c_0$ & $c_1$ & $c_2$ & $c_3$ & $c_4$ \\ \hline 
0.1892 & -0.1659 & 1.5497 & 0.6219 & 112.7136
\end{tabular} .
\end{split}
\end{equation}
These values generate the match of lattice QCD data \cite{Borsanyi:2013bia,Bazavov:2014pvz,Bellwied:2015lba} as documented in figures 1 and 2 of \cite{Knaute:2017opk} for thermodynamics and susceptibilities.

The thermodynamic quantities are calculated as
\begin{align}
  T &= \lambda_T          \frac{1}{4\pi \phi_A^{1/(4 - \Delta)} \sqrt{h_0^{\infty}}} , \quad
  s = \lambda_s          \frac{2 \pi}{\phi_A^{3/(4 - \Delta)}} , \\
  \mu &= \lambda_\mu  \frac{\Phi_0^{\infty}}{\phi_A^{1/(4 - \Delta)} \sqrt{h_0^\infty}} , \quad
  n = \lambda_n          \frac{f(\phi_0) \Phi_1}{2 f(0) \phi_A^{3/(4 - \Delta)}} , 
\end{align}
where the coefficients are extracted from a fit of the numerical solutions of $h(r), A(r), \Phi(r)$ and $\phi(r)$ to the ultraviolet boundary expansions \cite{DeWolfe:2010he}: 
$h(r) = h_0^{\infty} + \ldots$, 
$A(r) = \alpha(r) + \ldots$,
$\Phi(r) = \Phi_0^{\infty} + \Phi_2^{\infty} e^{-2 \alpha(r)} + \ldots$, and
$\phi(r) = \phi_A e^{ - (4 - \Delta) \alpha(r)}  + \phi_B e^{- \Delta \alpha(r)} + \ldots$ . 
Here, $\alpha(r) \equiv \frac{r}{L\sqrt{h_0^\infty}} + A_0^\infty$ and the scaling dimension of the field theory operator dual to $\phi$ follows from the horizon expansion of the potential $L^2 V(\phi) = - 12 + \frac 12 [\Delta(\Delta - 4)] \phi^2 + \ldots$ for $\phi \rightarrow 0$, implying $\Delta=2(1+\sqrt{1-3 a_1})$.
The dimensional scaling factors $\lambda_{T,s,\mu,n}$ restore physical units after setting $\kappa_5=L=1$ and satisfy $\lambda_T = \lambda_\mu := 1/L = \unit[1148.07]{MeV}$ and $\lambda_s = \lambda_n := 1/\kappa_5^2 = (\unit[513.01]{MeV})^3$.
}

\newpage 
\bibliographystyle{jk_ref_layout_noTitle}
\bibliography{literature}

\end{document}